\documentclass[aps, prd, twocolumn, nofootinbib, superscriptaddress, amsmath,amssymb]{revtex4}

\usepackage{graphicx}
\usepackage{epstopdf}
\usepackage{dcolumn}
\usepackage{bm}
\usepackage{amsmath, amssymb}
\usepackage[normalem]{ulem}
\usepackage{blindtext}
\usepackage{subfigure}
\usepackage{color}
\usepackage{multirow}

\newcommand{\beq}{\begin{eqnarray}}
\newcommand{\eeq}{\end{eqnarray}}
\newcommand{\tr}{\ensuremath{\mathrm{Tr}}}

\begin{document}

\title{Effects of a strong magnetic field on the QCD flux tube. }

\author{Claudio~Bonati}
\email{claudio.bonati@df.unipi.it}
\affiliation{Dipartimento di Fisica dell'Universit\`a di Pisa, Largo Pontecorvo 3, I-56127 Pisa, Italy}
\affiliation{INFN - Sezione di Pisa, Largo Pontecorvo 3, I-56127 Pisa, Italy}

\author{Salvatore~Cal\`i}
\email{scali@uni-wuppertal.de}
\affiliation{University of Cyprus, P.O. Box 20537, 1678 Nicosia, Cyprus}
\affiliation{University of Wuppertal, Gau{\ss}str. 20, 42119 Wuppertal, Germany}

\author{Massimo~D'Elia}
\email{massimo.delia@unipi.it}
\affiliation{Dipartimento di Fisica dell'Universit\`a di Pisa, Largo Pontecorvo 3, I-56127 Pisa, Italy}
\affiliation{INFN - Sezione di Pisa, Largo Pontecorvo 3, I-56127 Pisa, Italy}

\author{Michele~Mesiti}
\email{michele.mesiti@swansea.ac.uk}
\affiliation{Swansea University, Singleton Park, Swansea SA2 8PP, Wales, United Kingdom}

\author{Francesco~Negro}
\email{fnegro@pi.infn.it}
\affiliation{INFN - Sezione di Pisa, Largo Pontecorvo 3, I-56127 Pisa, Italy}

\author{Andrea~Rucci}
\email{andrea.rucci@pi.infn.it}
\affiliation{Dipartimento di Fisica dell'Universit\`a di Pisa, Largo Pontecorvo 3, I-56127 Pisa, Italy}
\affiliation{INFN - Sezione di Pisa, Largo Pontecorvo 3, I-56127 Pisa, Italy}

\author{Francesco~Sanfilippo}
\email{francesco.sanfilippo@roma3.infn.it}
\affiliation{INFN - Sezione di Roma Tre, Via della Vasca Navale 84, I-00146 Roma, Italy}

\date{\today}

\begin{abstract}
In this work we investigate the effect of an external magnetic field $B$ on the
shape of flux tubes in QCD by means of lattice simulations, performed with
$N_f=2+1$ flavors of stout improved dynamical staggered quarks with physical
masses.  After having discussed some difficulties in the practical definition
of the flux tube at $B=0$, we show that these ambiguities do not affect the
determination of the flux tube modifications induced by the magnetic field.
Different results are obtained depending on the relative orientations of the
flux tube and of the magnetic field: they confirm that the magnetic field acts
as transverse confinement catalyser and longitudinal confinement inhibitor;
moreover, the flux tube itself loses its axial symmetry when it is not directed
along the  magnetic background.
\end{abstract}


\maketitle

\section{Introduction}
\label{intro}

Despite the fact that a formal proof of color confinement in Quantum
Chromodynamics (QCD) is still lacking, Lattice QCD simulations provided an
overwhelming amount of numerical evidence that color confinement is encoded in
the QCD Lagrangian. Particularly important in this respect was the observation
\cite{Fukugita:1983du, Flower:1985gs, Wosiek:1987kx, Sommer:1987uz,
Bali:1994de, Haymaker:1994fm, Cea:1995zt} that the field generated by two
opposite static color sources is not a dipole field like in quantum
electrodynamics: most of the field energy density is concentrated in a linear
structure that connects the two static sources. The features of this linear
structure closely resemble the ones of the flux tubes experimentally observed
in type II superconductors, and for this reason the linear structure was called
color flux tube.  Analogous linear networks are observed when three or more
static color sources are used \cite{Okiharu:2003vt, Bissey:2006bz,
Bicudo:2011hk, Bakry:2014gea}.
 
The existence of the color flux tube provides an intuitive explanation for the
linearly rising potential between two opposite static color sources: the slope
of the potential (the string tension $\sigma$) is nothing but the energy
density per unit length of the flux tube.  As a consequence the study of the
color flux tube imposed itself as a tool to investigate the origin of the
confining potential in QCD in a way that is independent of the details of the
confining mechanism, even though the very idea of flux tube emerges very
naturally within the dual superconductor scenario for color confinement
\cite{tHooft:1975yol, Mandelstam:1974pi}.

The purpose of this paper is to provide a first Lattice QCD investigation of
flux tubes in the presence of a magnetic background field. Various lattice
studies have shown that such an external magnetic field has
a strong influence on the confining properties of QCD \cite{Bonati:2014ksa,
Bonati:2016kxj, Bonati:2017uvz, DElia:2015eey}, with the string tension in the
direction parallel to the magnetic field that is strongly reduced and which,
for large enough magnetic fields, could even disappear. Magnetic field induced
anisotropies could play a relevant role, especially at the level of heavy quark
phenomenology~\cite{Alford:2013jva, Giataganas:2013hwa, Giataganas:2013zaa,
Dudal:2014jfa, Cho:2014loa, Bonati:2015dka, Suzuki:2016kcs, Finazzo:2016mhm,
Yoshida:2016xgm, Suzuki:2016fof, Iwasaki:2018pby,Iwasaki:2018czv}. A possible
interpretation of such results can be found in various model
computations~\cite{Miransky:2002rp, Galilo:2011nh, Giataganas:2012zy,
Ferrer:2014qka, Rougemont:2014efa, Chernodub:2014uua, Miransky:2015ava,
Simonov:2015yka, Endrodi:2015oba, Schafer:2015wja, Dudal:2016joz,
Hasan:2017fmf, Giataganas:2018uuw,  Andreichikov:2018wrc}, and
looking at the flux tube provides a way to achieve a better comprehension of
the specific way in which the magnetic field influences the confining
properties of QCD.

\emph{A priori} various phenomena could indeed take place: the magnetic field
could change the strength of the color field within the flux tube, but it could
also modify the shape of the flux tube itself, which could even become
anisotropic and lose its axial symmetry when the quark-antiquark separation is
not collinear with the magnetic field. Even if the tube profile remains
cylindrical, one can look at the characteristic lengths that characterize the
flux tube and inquire how they are modified by the presence of an external
magnetic field.  

After the introduction in Sec.~\ref{setup} of the numerical setup adopted and
of the observables used, a preliminary part of our study is dedicated to the
investigation of the flux tube at zero magnetic field
(Sec.~\ref{resultszerofield}). Indeed, most studies in the literature have been
performed in pure gauge theories and only recently results for full QCD
appeared \cite{Cea:2017ocq}.  Therefore a study of the flux tube properties in
$N_f = 2+1$ QCD at the physical point, using a discretization different from
the one adopted in Ref.~\cite{Cea:2017ocq}, is interesting by itself. 

This preliminary part will give us the opportunity of discussing some
ambiguities related to the flux tube definition, that are associated with the
smoothing procedure adopted to improve the signal to noise ratio.  Remarkably
these ambiguities are much less severe (and indeed practically absent) if one
is interested only in modifications of the flux tube induced by the magnetic
field. This will be shown in Sec.~\ref{resultsnonzerofield}, where the main
results of this paper will be presented\footnote{Preliminary results have been
presented at the $35^{\mathrm{th}}$ International Symposium on Lattice Field
Theory (Lattice 2017) \cite{Bonati:2017anb}.}.  Finally, in Sec.~\ref{concl} we
report our conclusions.

\section{Numerical Setup} \label{setup}

\subsection{Lattice discretization of $N_f = 2+1$ QCD with a magnetic
background}

In this work we simulate $2+1$ flavour QCD making use of the stout improved rooted
staggered fermion discretization and the Symanzik tree-level improved gauge
action. More explicitly, the partition function is written as 
\begin{equation}\label{eq:partfunc}
Z(B) = \int \!\mathcal{D}U \,e^{-S_{Y\!M}}
\!\!\!\!\prod_{f=u,\,d,\,s} \!\!\!
\det{({D^{f}_{\textnormal{st}}[B]})^{1/4}}\ ,
\end{equation}
where $\mathcal{D}U$ stands for the product of the $SU(3)$ Haar measure of all
the links of the lattice. The gauge action $S_{Y\!M}$ is given
by~\cite{Weisz:1982zw, Curci:1983an}
\begin{equation}\label{eq:tlsyact}
S_{Y\!M}= - \frac{\beta}{3}\sum_{i, \mu \neq \nu} \left(
\frac{5}{6} P^{1\!\times \! 1}_{i;\,\mu\nu} - \frac{1}{12}
P^{1\!\times \! 2}_{i;\,\mu\nu} \right)\ ,
\end{equation}
where  the symbols $P^{1\!\times \! 1}_{i;\,\mu\nu}$ and $P^{1\!\times \!
2}_{i;\,\mu\nu}$ denote the real part of the trace of $1\!\times \!
1$ and $1\!\times \!2$ Wilson loops. The staggered Dirac matrix is
\begin{equation}\label{eq:rmmatrix}
\begin{aligned}
(D^f_{\textnormal{st}})_{i,\,j} =\ & am_f
  \delta_{i,\,j}+\!\!\sum_{\nu=1}^{4}\frac{\eta_{i;\,\nu}}{2}
  \left(u^f_{i;\,\nu}U^{(2)}_{i;\,\nu}\delta_{i,j-\hat{\nu}} \right. \\ 
  &-\left. u^{f*}_{i-\hat\nu;\,\nu}U^{(2)\dagger}_{i-\hat\nu;\,\nu}\delta_{i,j+\hat\nu}
  \right)\ , 
\end{aligned}
\end{equation}
where the $\eta_{i;\,\nu}$s are the usual staggered phases, $U^{(2)}_{i;\,\mu}$
stands for the two times stout-smeared link in position $i$ and direction $\mu$
\cite{Morningstar:2003gk} (with $\rho=0.15$ as isotropic smearing parameter)
and $u^f_{i;\,\mu}$ is the abelian field parallel transporter. 

The transporters corresponding to a uniform magnetic field $B_z$ directed
along $\hat{z}$ can be written as
\begin{eqnarray}\label{eq:bfield}
u^f_{i;\,y}=e^{i a^2 q_f B_{z} i_x} \ , \quad
{u^f_{i;\,x}|}_{i_x=N_x}=e^{-ia^2 q_f N_x B_z i_y}\, ,
\end{eqnarray}
where $q_f$ is the quark charge and all the other abelian link variables are
set to 1 ($N_k$ is the lattice extent in the direction $\hat{k}$ and $1\le
i_k\le N_k$).  However in this expression the value of $B_z$ cannot be
arbitrary: for Eq.~\eqref{eq:bfield} to describe a uniform magnetic field on a
lattice with periodic boundary conditions, the value $B_z$ has to satisfy the
quantization condition \cite{tHooft:1979rtg, Damgaard:1988hh, AlHashimi:2008hr}
\begin{equation}\label{bquant}
\frac{e}{3}B_z={2 \pi b}/{(a^2 N_x N_y)}\ ,
\end{equation}
where $b$ is an integer number. 

Let us stress that the magnetic field is
external: abelian transporters $u^{f}_{i;\,\mu}$ are not updated, so quarks
interact with the external magnetic field but they do not back-react on it. In
this way we are neglecting the direct quark-quark electromagnetic interactions,
while we are properly taking into account the effect of the external field on
the quark loops.

Bare parameters have been chosen in such a way that simulations stay on a line
of constant physics with physical values of the quark masses.  Since the
lattice spacing is independent of the magnetic field \cite{Bali:2011qj}, we
could use for this purpose the values reported in Refs.~\cite{Aoki:2009sc,
Borsanyi:2010cj, Borsanyi:2013bia}.  Gauge configurations have been sampled by
using the Rational Hybrid Monte-Carlo (RHMC) \cite{Clark:2004cp, Clark:2006fx,
Clark:2006wp} algorithm; simulation parameters and details are reported in
Table~\ref{tab:sim_details}.

\begin{table}[b]
\centering
\begin{tabular}{|l|l|c|l|l|l|c|}
\hline
$\beta$ & $m_l$ & Lattice & $a$ [fm] & $b$ & $eB$ [GeV$^2$] & $N_{\mathrm{confs}}$ \\
\hline
3.7500 & 0.001787 & \rule{0mm}{3mm}$40^4$ & 0.1249 & 0 & \quad 0 &  51 \\
\hline
\multirow{6}{*}{3.8500} & \multirow{6}{*}{0.001400} & \multirow{6}{*}{$48^3\times 96$} & \multirow{6}{*}{0.0989} & 0  & \quad 0    & 19 \\
                   &                    &                                  &                         & 8  & \quad 0.26 & 28 \\
                   &                    &                                  &                         & 24 & \quad 0.78 & 43 \\
                   &                    &                                  &                         & 32 & \quad 1.04 & 20 \\
                   &                    &                                  &                         & 64 & \quad 2.08 & 16 \\
                   &                    &                                  &                         & 96 & \quad 3.12 & 20 \\
\hline
\multirow{3}{*}{3.8950} & \multirow{3}{*}{0.001274} & \multirow{3}{*}{$48^4$} & \multirow{3}{*}{0.0898} & 0  & \quad 0    & 23 \\
                   &                    &                         &                         & 16 & \quad 0.63 & 19 \\
                   &                    &                         &                         & 24 & \quad 0.94 & 13 \\
\hline
\multirow{3}{*}{3.9575} & \multirow{3}{*}{0.001130} & \multirow{3}{*}{$48^4$} & \multirow{3}{*}{0.0796} & 0  & \quad 0    & 22 \\
                   &                    &                         &                         & 16 & \quad 0.80 & 14 \\
                   &                    &                         &                         & 24 & \quad 1.20 & 10 \\
\hline
\end{tabular}
\caption{Simulation details: the bare coupling $\beta$, the bare light quark
mass $m_l$ ($m_s/m_l$ was always fixed to 28.15), the lattice size, the value
of the lattice spacing (with a systematic uncertainty of 2-3\%,
see~\cite{Aoki:2009sc, Borsanyi:2010cj, Borsanyi:2013bia}), the magnetic
quantum $b$, the magnetic field intensity $eB$ and the number of independent
configurations $N_{\mathrm{confs}}$ analyzed, separated by $10\div 25$
molecular dynamics trajectories.}\label{tab:sim_details}
\end{table}

\begin{figure}
\includegraphics*[width=0.75\columnwidth]{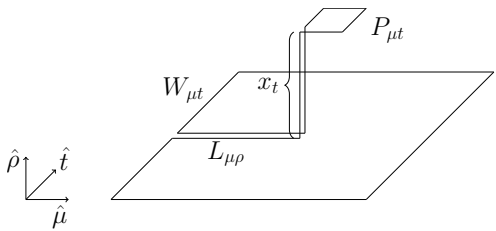}
\caption{Path used to define the first term of the connected observable
$\rho_{conn}$ in Eq.~\eqref{eq:conn_corr}. The orientations of the Wilson loop
and of the plaquette can in general be different, however in this paper we
focus on the longitudinal component of the chromoelectric field, in which case
$W$ and $P$ are parallel to each other, as in the figure.}
\label{fig_operator} 
\end{figure}

\subsection{Observables}

To study the color flux tube in a lattice simulation two basic ingredients are
needed: two static color sources of opposite charge, that are usually
introduced by means of a Wilson loop (or two Polyakov loops if the temperature
is non-vanishing), and a probe to investigate the field structure, that is
usually a plaquette, i.e. a Wilson loop of size $1\times 1$.  The field profile
around the static charges is extracted from the correlation of the plaquette
with the Wilson loop for several positions and orientations of the plaquette.

Two different practical implementations of this general idea exist in the
literature: in the first case \cite{Fukugita:1983du} one studies the
disconnected correlator between the Wilson loop and the plaquette
\begin{equation}\label{eq:disc_corr}
\rho_{disc}(W, P)=\frac{\langle \tr(W)\tr(P)\rangle}{\langle\tr(W)\rangle}-\langle \tr(P)\rangle\ ,
\end{equation} 
in the second case \cite{DiGiacomo:1989yp, DiGiacomo:1990hc} the connected
correlator is used instead 
\begin{equation}\label{eq:conn_corr}
\rho_{conn}(W,P)=\frac{\langle \tr(WLPL^{\dag})\rangle}{\langle\tr(W)\rangle}-
\frac{1}{N_c}\frac{\langle\tr(P)\tr(W)\rangle}{\langle\tr(W)\rangle}\ ,
\end{equation}
where $N_c$ is the number of color and $L$ is a parallel transporter that
connects the Wilson loop and the plaquette (which is often called ``Schwinger
line''), see Fig.~\ref{fig_operator} for a pictorial representation.
Note that in both cases, given that in the present
case the electromagnetic field is not dynamical, a possible 
$U(1)$ charge of the static pair is irrelevant, since it would 
just add a phase to the Wilson loop which cancels in the ratios, while
the plaquette operator is by definition limited to the $SU(3)$ part,
since we are interested in the modifications of the non-Abelian fields
in the flux tube induced by the external magnetic background.

These two lattice implementations are not completely equivalent, as they probe
different quantities: in the \emph{naive} continuum limit the disconnected
correlator
$\rho_{disc}$ reduces to 
\begin{equation}\label{eq:disc_cont}
\begin{aligned}
\rho_{disc}^{\mu\nu}&\simeq -a^4g_0^2\left(\frac{\langle\tr(W)\tr[F_{\mu\nu}^2]\rangle}{\langle\tr(W)\rangle}-
\langle\tr[F_{\mu\nu}^2]\rangle\right)=\\
&=-a^4g_0^2\Big(\langle \tr[F_{\mu\nu}^2(x)]\rangle_{Q\bar{Q}}-\langle \tr[F_{\mu\nu}^2(x)]\rangle_0
\Big)\ , 
\end{aligned}
\end{equation} 
where $\mu$ and $\nu$ identify the plaquette orientation (no sum on $\mu,\nu$
is intended in the r.h.s), $F_{\mu\nu}$ is the continuum euclidean field
strength, $a$ is the lattice spacing and $g_0$ is the bare coupling constant.
The subscripts $Q\bar{Q}$ and $0$ are used in the previous expression to denote
the cases in which two opposite static color charges are present in the
background or not.  When using the connected correlator $\rho_{conn}$,
the plaquette operator appears in the same trace of the Wilson loop operator, so
in the \emph{naive} continuum limit the term that is linear in the nonabelian
field strength at the position of the probe gives the leading contribution
\begin{equation}\label{eq:conn_cont}
\rho_{conn}^{\mu\nu}\simeq a^2g_0\frac{\langle\tr[iWLF_{\mu\nu}L^{\dag}]\rangle}{\langle \tr(W)\rangle}\ ,
\end{equation}
an expression that in the literature is often denoted, for the sake of 
simplicity but with a clear abuse of notation, by $a^2g_0\langle
F_{\mu\nu}\rangle_{Q\bar{Q}}$. 	 

While $\rho_{conn}$ and $\rho_{disc}$ display similar behaviors as a function
of the transverse displacement $x_t$, the connected correlator has a
significantly larger signal to noise ratio.  This is related to the fact that
using $\rho_{conn}$ we access the field strength and not its square; as a
consequence $\rho_{conn}$ is much less sensitive to the fluctuations, which
also means that it is expected to be more stable under smoothing of the gauge
fields. In some cases one is interested precisely in fluctuation effects (e.g.
in the study of the fluctuation-induced broadening of the flux tube) and it is
then mandatory to use $\rho_{disc}$, which in pure gauge theories can be
precisely estimated using standard noise reduction techniques
\cite{Allais:2008bk, Gliozzi:2010zv, Gliozzi:2010jh, Cardoso:2013lla,
Amado:2013rja}; if this is not the case and large statistics are not available
$\rho_{conn}$ is a more convenient choice \cite{Cardaci:2010tb,Cea:2012qw,
Cea:2014uja, Cea:2015wjd, Cea:2017ocq}. For these reasons and to directly
compare with the recent results \cite{Cea:2017ocq} we used the connected
correlator $\rho_{conn}$ defined in Eq.~\eqref{eq:conn_corr}.

The geometry of the connected correlator $\rho_{conn}$ is depicted in
Fig.~\ref{fig_operator}: the Schwinger line $L$ is attached to the square
Wilson loop $W$ in the midpoint of its temporal extent, it reaches half the
distance between the static color sources and then it moves $x_t$ lattice
spacings in a direction orthogonal to the Wilson loop plane.  Since
$\rho_{conn}$ is a purely gluonic observable, the magnetic field can affect its
value only by loop (sea) effects. In particular, to study different values of
$B$, different sets of configurations have to be generated. When evaluating
$\rho_{conn}$ it is important to realize that the external magnetic field
breaks the lattice octahedral symmetry and Wilson loops oriented along
different directions in general will not be equivalent; analogously, the two
directions that are orthogonal to the plane of the Wilson loop will not be
equivalent for generic magnetic field orientations (see later discussion).

In order to reduce the UV noise and improve the signal to noise ratio we
adopted, as usual, smearing.  With the aim of simplifying the comparison with
previous results in the literature, we chose  the same smearing procedure
adopted in Ref.~\cite{Cea:2017ocq}: a single HYP smearing step
\cite{Hasenfratz:2001hp} has been applied to all the temporal links, with
parameters $(\alpha_1,\alpha_2, \alpha_3)=(1.0,0.5, 0.5)$, then several APE
smearing steps \cite{Albanese:1987ds} have been applied to the spatial links,
according to the definition
\begin{equation}\label{eq:APE}
U_{\mu}^{APE}(x)=\mathrm{Proj}_{SU(3)}\Big( U_{\mu}(x)+\alpha_{APE}S_{\mu}(x)\Big)\ ,
\end{equation}
where $\alpha_{APE}$ was fixed to $1/6$ and $S_{\mu}(x)$ is the sum of all the
spatial staples associated to the spatial link $U_{\mu}(x)$.

\section{Results for $N_f = 2+1$ QCD at $B = 0$}
\label{resultszerofield}

In order to investigate the profile of the flux tube we measured the
longitudinal (i.e.~directed along the flux tube) component of the
chromoelectric field $E_{l}$, since all previous studies 
showed this component of the field strength to be the dominant one. If we
denote by $\hat{\mu}$ the axis of the relative separation between the two color
charges, the longitudinal chromoelectric field is given by 
\begin{equation}
E_l(d, x_t) = \frac{1}{a^2}\sqrt{\frac{\beta}{6}}\rho_{conn}^{(t,\mu)}(d, x_t)\ ,
\end{equation}
where $d$ is the distance between the charges, $x_t$ is the transverse distance
at which the flux profile is probed and $(t,\mu)$ is the plaquette
orientation. 

\begin{figure}[b]
\includegraphics*[width=\columnwidth]{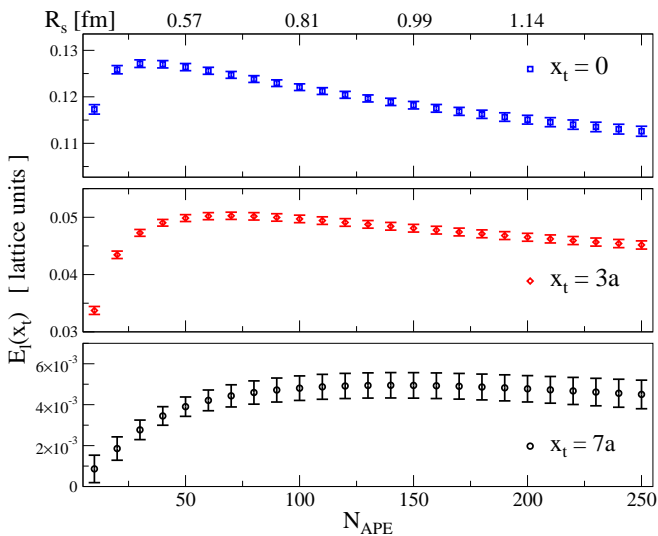}
\caption{Dependence on the number of APE smearing steps of longitudinal
chromoelectric field for $d\simeq 0.7$~fm and three $x_t$ values, measured on a
$48^3 \times 96$ lattice with lattice spacing $a\simeq 0.0989$~fm ($R_s$ is the smoothing
radius in physical units defined in Eq.~\eqref{eq:R_s}).}
\label{flux_dep_APE}
\end{figure}

Since smearing is used 
to reduce the UV noise, it is important to study
the dependence of the results on the amount of smoothing adopted, in order to
asses the reliability of the results.  As an example, in
Fig.~\ref{flux_dep_APE} we report the values obtained for $E_l(d, x_t)$ using a
$48^3 \times 96$ lattice, with $a\simeq 0.0989~\mathrm{fm}$ and $d\simeq
0.7~\mathrm{fm}$: results are shown, as function of the number of APE smearing
steps $N_{APE}$, for three different values of the transverse separation $x_t$.
It is clear that a non-trivial dependence on $N_{\rm APE}$ is present 
and a prescription is needed to fix the value of $\rho_{conn}$. 

In Refs.~\cite{Cea:2012qw,Cea:2014uja, Cea:2015wjd, Cea:2017ocq} the
prescription adopted was to take the value at the plateau (or at the maximum)
which is reached after some smearing steps; this is analogous to what has been
done in the literature for similar quantities, like the gauge-invariant field
strength correlators~\cite{FC_0,FC_1,DElia:2015eey}.  
This prescription implies that the field value has to
be taken after different numbers of smearing steps for different values of
$x_t$, since the plateau (or maximum) is reached for larger values of
$N_{APE}$ when increasing $x_t$, as can be seen in Fig.~\ref{flux_dep_APE}.

In this study we explore also a different prescription, in which all field
strength values at different $x_t$ (and also at different values of the lattice
spacing $a$) are measured keeping constant the smearing radius $R_s$ in
physical units. The continuum limit is then taken at fixed $R_s$ and results
obtained using different smoothing radii can be compared among them.  Such a
prescription is similar to the one adopted to compute renormalized observables
using the gradient flow as a regulator (see
e.g.~\cite{Luscher:2010iy}), and the similarity is even more striking in view
of the practical equivalence between the smoothing techniques
\cite{Bonati:2014tqa, Alexandrou:2015yba, Alexandrou:2017hqw, Berg:2016wfw}. 

We fixed the value of the smearing radius $R_s$ in physical units according to the
following relation (obtained in Ref.~\cite{Alexandrou:2017hqw})
\begin{equation}\label{eq:R_s}
R_s=a\sqrt{\frac{8 \alpha_{APE}}{1+6\alpha_{APE}}N_{APE}} \, .
\end{equation}
This expression slightly differs from the one reported in
\cite{Alexandrou:2017hqw}, since in the present work we adopt a
normalization of the parameter entering the APE smearing that
is different from the one used in the original derivation:
\begin{equation}
\alpha_{APE}^{\mathrm{[77]}}=\frac{\alpha_{APE}}{1+\alpha_{APE}}\ .
\end{equation}
A further difference is that in this study we use only the spatial staples in
the APE smearing and we update only the spatial links, while in
Ref.~\cite{Alexandrou:2017hqw} a four dimensional smearing was studied. This would
results in a further multiplicative factor in Eq.~\eqref{eq:R_s}, independent
of $N_{APE}$ and $a$. Since our aim is to perform the continuum extrapolation
at fixed $R_s$, this multiplicative factor is practically irrelevant and in the
following we will just use Eq.~\eqref{eq:R_s}.

\begin{figure}
\includegraphics*[width=\columnwidth]{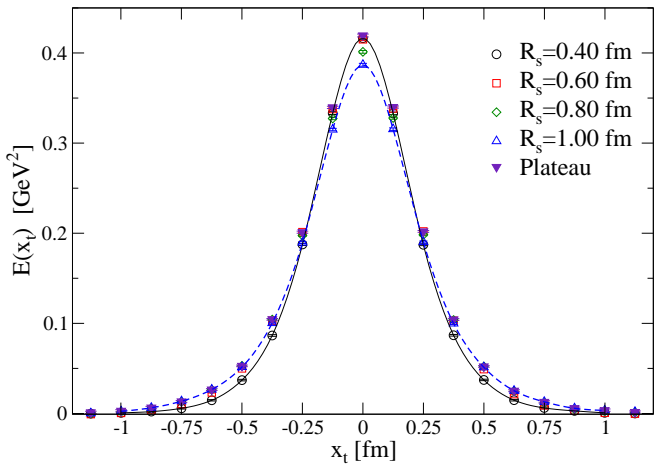}
\caption{Flux tube shapes obtained using different values of the smearing
parameter $R_s$ and using the plateau method. The distance between the two
opposite static color charges was fixed to $d=0.7~\mathrm{fm}$ and measures
were performed on a $40^4$ lattice with $a\simeq 0.1249~\mathrm{fm}$.
Continuous lines are cubic spline interpolations and are shown for the cases
$R_s\simeq 0.4$~fm and $R_s\simeq 1.0$~fm in order to guide the eye.}
\label{compare_Rs_plateau}
\end{figure}

In Fig.~\ref{compare_Rs_plateau} we show some results for the flux tube profile
extracted from simulations performed at lattice spacing $a=0.1249~\mathrm{fm}$
(on a $40^4$ lattice) using a Wilson loop of physical size $d\simeq
0.7~\mathrm{fm}$: different symbols correspond to results obtained using
different values of the smoothing radius $R_s$ and values extracted from the
plateaux are also shown for comparison. From Fig.~\ref{compare_Rs_plateau} it
can be seen that not only the absolute scale of the flux tube depends on $R_s$
but also its shape. Similar behaviors are observed for all the values of
$a$ and $d$ explored in this work.

Two different strategies can be adopted in order to keep the size of the Wilson
loop constant in physical units while changing the lattice spacing.  An
approach consists in fixing \emph{a priori} the extent of the Wilson loop in
lattice units and the value of lattice spacing, imposing the constraint of
constant physical size. A different possibility is to perform measures, for
each value of the lattice spacing, using several Wilson loop extents, in order
to be able to interpolate the results on a wide range of sizes. This second
possibility requires some more care during the analysis, to check for possible
systematics induced by the interpolation procedure; on the other hand it is
much more flexible, since one can \emph{a posteriori} decide the optimal size
to be used in order to have small systematics and good signal to noise ratio.
For this reason we adopted the second possibility. 

\begin{figure}[b]
\includegraphics*[width=\columnwidth]{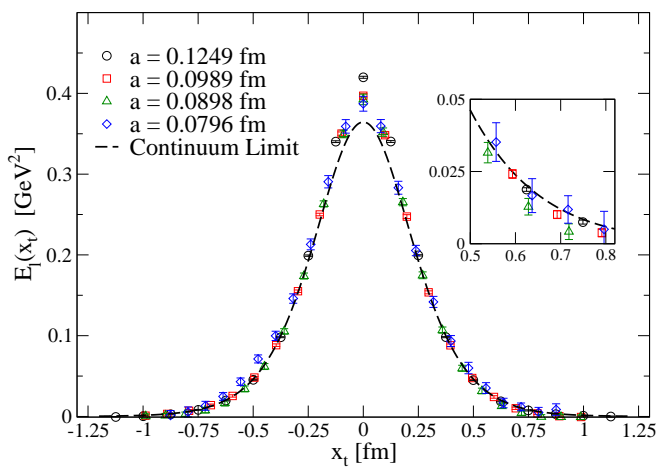}
\caption{Dependence of the flux tube profile on the lattice spacing at fixed
distance between the color sources ($d\simeq 0.7$~fm) and fixed smoothing
radius ($R_s\simeq 0.5$~fm).  The dashed line represents the continuum limit
performed according to the Clem ansatz (see text).}
\label{fig:several_a_fixed_Rs}
\end{figure}

We verified that the interpolation in the smearing radius is stable (i.e.
independent of the interpolating function adopted) to a high level of accuracy,
as could have been expected given the fact that the dependence on $R_s$ is
quite smooth. More care is needed for the interpolation in the distance $d$
between the static color charges, since we have less data points available: for
the coarsest lattice we measured Wilson loop from $4\times 4$ up to $7\times
7$, reaching $11\times 11$ for the finest lattice. However, starting from
$d\gtrsim 0.6$~fm, results are independent of the order of the spline
interpolation adopted (linear, quadratic and cubic splines were tested).  Using
these interpolations we can compare the results obtained at different lattice
spacings and, as an example, in Fig.~\ref{fig:several_a_fixed_Rs} we show the
outcome of this analysis for $d\simeq 0.7$~fm and $R_s\simeq 0.5$~fm: finite
cut-off corrections seem to be small and more visible for small values of
$x_t$; similar considerations apply to results obtained for different values of
$R_s$ and for the plateau method.

In order to perform the continuum limit of our results in a model independent
way, data at all lattice spacings should be available for each value of $x_t$.
As one can see from Fig.~\ref{fig:several_a_fixed_Rs}, because of the discrete
nature of the lattice distances this is not the case.  In principle, we could
think of adopting the same strategy used above and interpolate, for each
lattice spacing, results obtained at different values of $x_t$. However, in
this case we find it more convenient to make use of a well known ansatz for the
flux tube profile.

In particular, a parametrization that was shown in previous studies
\cite{Cea:2012qw, Cea:2014uja, Cea:2015wjd, Cea:2017ocq} to well describe the
data for the longitudinal component of the electric field is the Clem form
\begin{equation}\label{eq:clem_fit}
E_l(x_t)=\frac{\phi}{2\pi}\frac{\mu^2}{\alpha}\frac{K_0\big(\sqrt{\mu^2 x_t^2+\alpha^2}~\big)}{K_1(\alpha)}\ ,
\end{equation} 
where $\phi, \alpha$ and $\mu$ are fit parameters and $K_0$, $K_1$ are modified
Bessel functions of the second kind. This parametrization of the longitudinal
(chromo-)electric field is inspired by a similar parametrization of the
longitudinal magnetic field around a vortex in type II superconductors, that
was obtained in Ref.~\cite{Clem} by variationally improving an \emph{ansatz}
solution of the Ginzburg-Landau equations.  We checked that
Eq.~\eqref{eq:clem_fit} is consistent with the observed flux tube profiles for
all the values of the lattice spacing, of the smoothing radius and of the
quark-antiquark distance explored in the present work.

For the purpose of performing the continuum limit, Eq.~\eqref{eq:clem_fit} is
just a reasonably simple functional form that well describes data using three
parameters. From a broader perspective, however, the fact that
Eq.~\eqref{eq:clem_fit} well describes lattice data supports the dual
superconductivity picture of confinement. In this picture condensation of the
chromomagnetic degrees of freedom is expected to happen in the vacuum,
confinement is the chromoelectic analogue of the standard Meissner effect and,
in complete analogy with ordinary superconductors, the vacuum is characterized
by two length scales: the penetration length $\lambda$ and the coherence length
$\xi$.  These scales are related to the parameters entering
Eq.~\eqref{eq:clem_fit} by the relations (see Ref.~\cite{Clem})
\begin{equation}\label{eq:lambda_kappa}
\mu=\frac{1}{\lambda}\ , \quad
\kappa=\frac{\sqrt{2}}{\alpha}\sqrt{1-\frac{K_0^2(\alpha)}{K_1^2(\alpha)}}\ ,
\end{equation}
where $\kappa=\lambda/\xi$ is the Ginzburg-Landau parameter, whose value
discriminates between type I superconductors, corresponding to
$\kappa<1/\sqrt{2}$, and type II superconductors, for which $\kappa >
1/\sqrt{2}$ (see, e.g., Ref.~\cite{Tinkham}).

To investigate the lattice spacing dependence of the results and to extract their
continuum limit we used Eq.~\eqref{eq:clem_fit} with $a$-dependent parameters.
Since in our lattice discretization the leading lattice artefacts are
$\mathcal{O}(a^2)$, we performed a global fit to all the data corresponding to
fixed values of $d$ (quark-antiquark separation) and $R_s$ (smoothing radius),
using the functional form in Eq.~\eqref{eq:clem_fit} with the substitutions
\begin{equation}
\begin{aligned}
\phi &\to\phi_0+a^2\phi_1\\
\mu  &\to \mu_0+a^2\mu_1\\
\alpha &\to \alpha_0+a^2\alpha_1 \ .
\end{aligned}
\end{equation}
Quantities denoted by the ``0'' subscript are the continuum values of 
$\phi$, $\mu$ and $\alpha$,  while quantities with subscript ``1''
parametrize lattice artifacts.

\begin{figure}
\includegraphics*[width=\columnwidth]{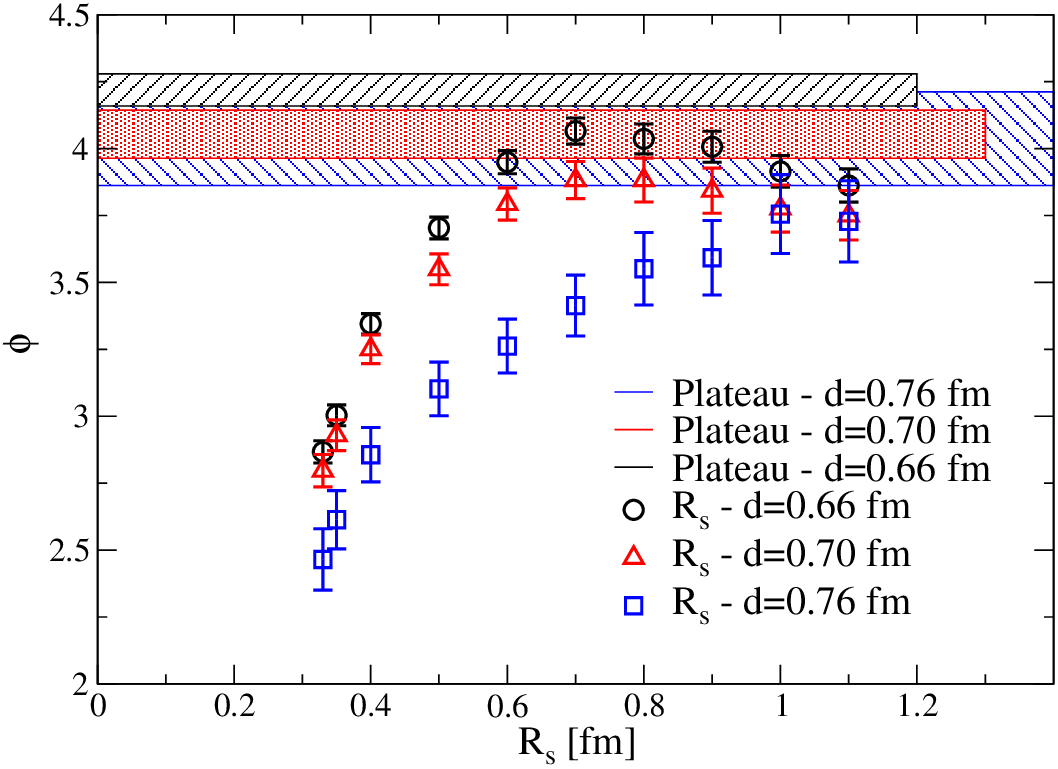}
\caption{Continuum limit of the parameter $\phi$ of Eq.~\eqref{eq:clem_fit}:
comparison of the extrapolations obtained via the fixed smearing
radius approach and the plateau method.}
\label{fig:phi_comparison}
\end{figure}

\begin{figure}
\includegraphics*[width=0.98\columnwidth]{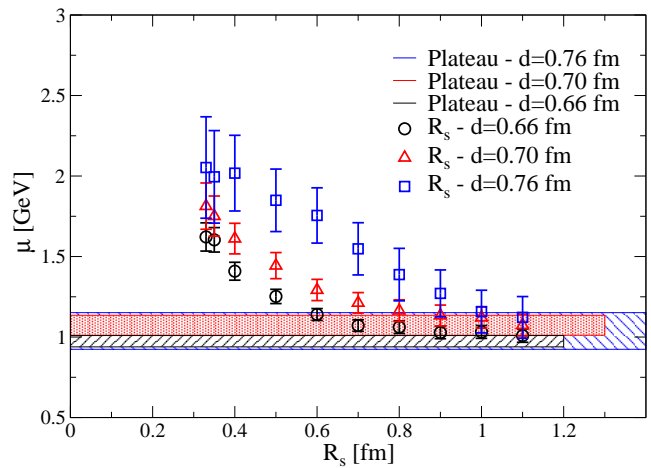}
\caption{Continuum limit of the parameter $\mu$ of Eq.~\eqref{eq:clem_fit}:
comparison of the extrapolations obtained via the fixed smearing
radius approach and of the plateau method.}
\label{fig:mu_comparison}
\end{figure}

\begin{figure}
\includegraphics*[width=0.97\columnwidth]{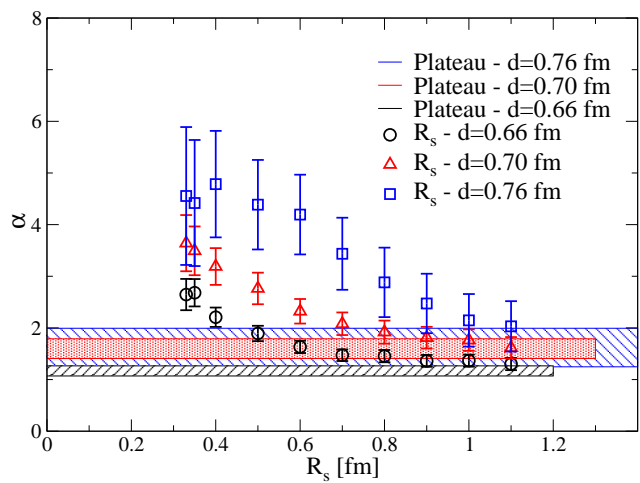}
\caption{Continuum limit of the parameter $\alpha$ of Eq.~\eqref{eq:clem_fit}:
comparison of the extrapolations obtained via the fixed smearing
radius approach and of the plateau method.}
\label{fig:alpha_comparison}
\end{figure}

\begin{figure}
\includegraphics*[width=\columnwidth]{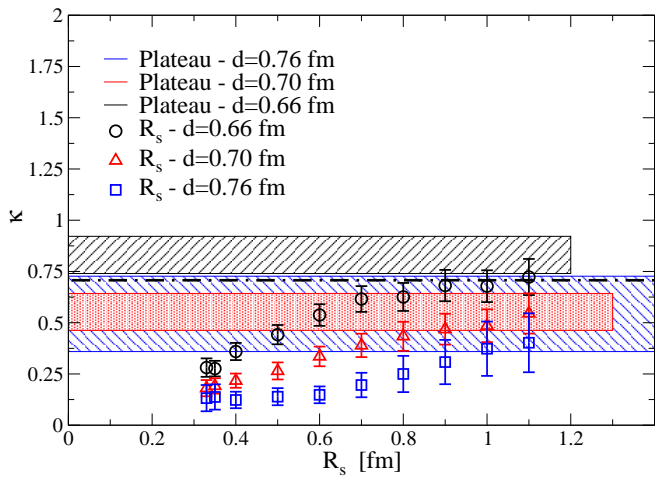}
\caption{Continuum limit of the Ginzburg-Landau parameter $\kappa$ (obtained
from Eq.~\eqref{eq:lambda_kappa}): comparison of the extrapolations obtained
via the fixed smearing radius approach and of the plateau method. The dotted-dashed
horizontal line corresponds to the critical value $1/\sqrt{2}$.}
\label{fig:kappa_comparison}
\end{figure}

The global fit works reasonably well for all explored values of $R_s$ as well
as for the plateau method. For instance, for the data in
Fig.~\ref{fig:several_a_fixed_Rs} we obtain a value of the $\chi^2/{\rm
d.o.f.}$ around 1.3; moreover the fit parameters are stable, within errors,
when one eliminates data at the coarsest lattice spacing from the fit.  The
continuum values of the parameters $\phi$, $\mu$ and $\alpha$ that are obtained
in this way are shown in Figs.~\ref{fig:phi_comparison},
\ref{fig:mu_comparison} and \ref{fig:alpha_comparison} for three different
values of the distance between the static color charges ($d$) and several
values of the smoothing radius ($R_s$). In all cases a sizable dependence of
the results on $d$ and $R_s$ can be seen. Results computed at fixed smoothing
radius converge, for large $R_s$, to the values extracted using the plateau
method, which displays only a weak dependence on the value of $d$ used.  Also
the values of derived quantities like the Ginzburg-Landau parameter $\kappa$
are dependent on the specific values of $d$ and $R_s$ used, as shown in
Fig.~\ref{fig:kappa_comparison}.  In principle, one could try an extrapolation
of continuum results to zero smearing radiuds, $R_s= 0$, however our present
accuracy does not permit to perform that reliably.

For all the combinations of $d$ and $R_s$ values studied we did not
found results incompatible with $\kappa \lesssim 1/\sqrt{2}$: only for $d\simeq
0.66\,\mathrm{fm}$, using the plateau method or large smoothing radii, results
for $\kappa$ larger than this critical value are obtained, which are however
compatible with it at one standard deviation. This behavior favors the
interpretation of the QCD vacuum as a type I superconductor, in accordance with
previous results in the literature~\cite{Cea:2017ocq}, however the strong
dependence of $\kappa$ on the values of $d$ and $R_s$ makes it very difficult
to draw firm conclusions on the actual value of $\kappa$ in QCD.

\section{Results in a magnetic background field}
\label{resultsnonzerofield}

The numerical results that will be presented in this section refer, unless
otherwise explicitly stated, to the $48^3\times 96$ lattice with lattice
spacing $a\simeq 0.0989~\mathrm{fm}$, that is the lattice size on which the
largest number of simulations with $B\neq 0$ have been performed (see
Table~\ref{tab:sim_details}). 

The presence of the external magnetic field explicitly breaks some of the
symmetries of the QCD Lagrangian, both in the continuum and on the lattice.
Particularly important for our purpose is the breaking of the rotation
symmetry: the details of the flux tube profile will depend on the relative
orientations of the magnetic field, the Wilson loop and the transverse
direction chosen to probe the chromoelectric field.

\begin{table}
\begin{tabular}{|c|c|c|}
\hline 
\parbox{1cm}{$\hat{\mu}$}  & \parbox{1cm}{$\hat{\rho}$} & class \\ \hline 
$\hat{z}$                  & $\hat{x}$                  & \multirow{2}{*}{L} \\ 
$\hat{z}$                  & $\hat{y}$                  & \\ \hline
$\hat{x}$                  & $\hat{y}$                  & \multirow{2}{*}{TT} \\ 
$\hat{y}$                  & $\hat{x}$                  & \\ \hline
$\hat{x}$                  & $\hat{z}$                  & \multirow{2}{*}{TL} \\ 
$\hat{y}$                  & $\hat{z}$                  & \\ \hline
\end{tabular}
\caption{Equivalence classes of flux tubes, identified by the relative
orientations of the magnetic field (always assumed to be directed along
$\hat{z}$), the Wilson loop (in the plane $(t,\mu)$) and the transverse
direction $\hat{\rho}$, see Fig.~\ref{fig_operator}.}\label{tab:eq_classes}
\end{table}

In this work the magnetic field will always be directed along one of the
lattice axes, which we can assume to be the $\hat{z}$ direction.  The plane of
the Wilson loop is identified by the couple of indices $(t,\mu)$, with $\mu \in
\{x, y, z\}$, and we denote by $\hat{\rho}$ the spatial direction, orthogonal
to the Wilson loop, along which the chromoelectric field is evaluated (see
Fig.~\ref{fig_operator}).  If the color sources are separated along the
magnetic field, i.e. $\hat{\mu}=\hat{z}$, the theory is invariant under
rotations around the $\hat{z}$ axis and the two possible choices
$\hat{\rho}=\hat{x}$ and $\hat{\rho}=\hat{y}$ of transverse direction are
equivalent. If instead $\hat{\mu}=\hat{x}$, the orthogonal directions
$\hat{\rho}=\hat{y}$ and $\hat{\rho}=\hat{z}$ are not equivalent, and an
analogous situation happens for $\hat{\mu}=\hat{y}$. It is however simple to
verify that the two combinations $\hat{\mu}=\hat{x}$, $\hat{\rho}=\hat{y}$ and
$\hat{\mu}=\hat{y}$, $\hat{\rho}=\hat{x}$ can be mapped into each other by
using a rotation along the $\hat{z}$ axes and a reflection with respect to a
plane containing the $\hat{z}$ axis, which are symmetry transformations also
when a non-vanishing magnetic field is present.  In a similar way it can be
shown that the choice $\hat{\mu}=\hat{x}$, $\hat{\rho}=\hat{z}$ is equivalent
to $\hat{\mu}=\hat{y}$, $\hat{\rho}=\hat{z}$. We thus have, under the residual
symmetry that is present when $B\neq 0$, three equivalence classes of flux
tubes, that will be denoted by the shorthand L, TT, TL (T stands for
transverse and L for longitudinal with respect to the magnetic field) and are
reported in Table~\ref{tab:eq_classes} for later reference.

\begin{figure}
\includegraphics*[width=\columnwidth]{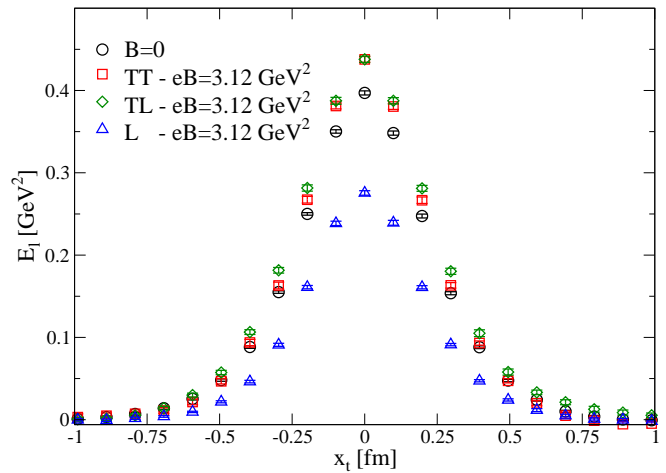}
\caption{Dependence of the flux tube profile on the orientation of the magnetic
field. Results obtained for $eB\simeq 3.12~\mathrm{GeV}^2$ ($d\simeq
0.7~\mathrm{fm}$, $R_s\simeq 0.5~\mathrm{fm}$) are compared with the ones
at vanishing magnetic field. Notation is defined in
Table~\ref{tab:eq_classes}. }
\label{fig:flux_B_dependence}
\end{figure}

The flux tube shapes obtained in the three inequivalent geometries of
Table~\ref{tab:eq_classes} are shown in Fig.~\ref{fig:flux_B_dependence} for
$eB\simeq 3.12~\mathrm{GeV}^2$ (using $R_s\simeq 0.5\,\mathrm{fm}$). The
intensity of the longitudinal electric field when the flux tube is directed
along the magnetic field is strongly reduced with respect to the case in which
it is transverse to it. Moreover, in the transverse case, the flux tube loses
its axial symmetry, as seen from the fact that the results for the cases TT and
TL are different from each other. Notice that, 
in all cases, the symmetry
under the transformation $x_t\to -x_t$ is preserved:
in principle, 
one could expect asymmetries when the magnetic field is orthogonal 
to the quark-antiquark separation, however that would imply
that the magnetic field induces a component of the chromoelectric field
parallel to it, and this is protected by the CP symmetry, which is 
not broken by the magnetic background.

\begin{figure}
\includegraphics*[width=\columnwidth]{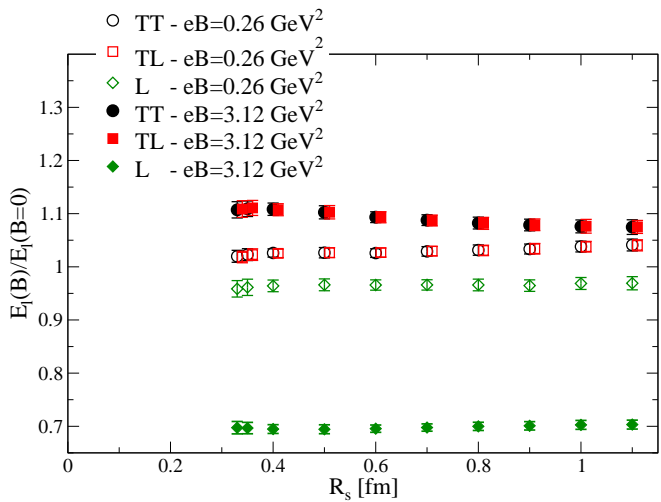}
\caption{Dependence on the smoothing radius $R_s$ of the ratios $E_l(d, x_t,
R_s, B)/E_l(d, x_t, R_s, B=0)$, computed for $eB\simeq 0.26~\mathrm{GeV}^2$ and
$eB\simeq 3.12~\mathrm{GeV}^2$, $d\simeq 0.7~\mathrm{fm}$ and $x_t\simeq 0~\mathrm{fm}$.}
\label{fig:profratio_B}
\end{figure}

\begin{figure}
\includegraphics*[width=\columnwidth]{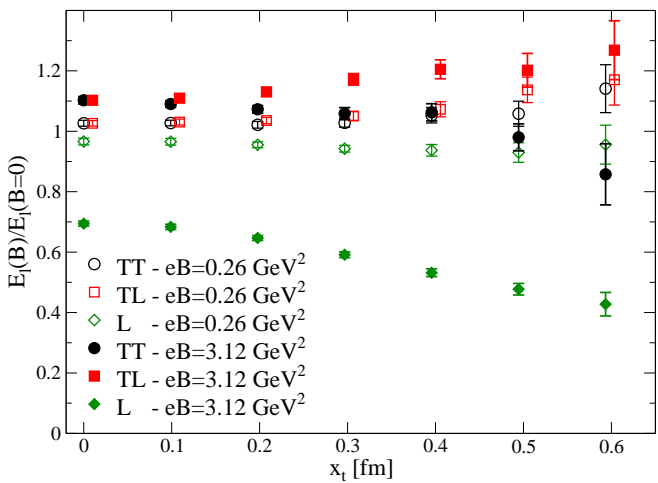}
\caption{Dependence of the flux tube profile on the magnetic field: values of $E_l(d,
x_t, B)/E_l(d, x_t, B=0)$ computed for $eB\simeq 0.26~\mathrm{GeV}^2$ and
$eB\simeq 3.12~\mathrm{GeV}^2$, with $d\simeq 0.7~\mathrm{fm}$.}
\label{fig:tube_ratio_B}
\end{figure}

Also in the presence of a non-vanishing magnetic field the details of the flux
tube strongly depend on the value of the smoothing radius $R_s$. However the
ratio of the chromoelectric fields with and without the magnetic field (i.e.
$E_l(d, x_t, R_s, B)/E_l(d, x_t, R_s, B=0)$) is remarkably insensitive to the
value of $R_s$. This is true for all the inequivalent classes of
Table~\ref{tab:eq_classes} and for all the values of the transverse distance
$x_t$ studied, with some examples shown in Fig.~\ref{fig:profratio_B}. This
means that we can study the effect of the magnetic field on the flux tube in an
unambiguous way and for this reason the dependence on $R_s$ will be dropped in
the following.

Fig.~\ref{fig:tube_ratio_B} shows the changes in the flux tube induced by the
magnetic field. The most striking effect that can be seen is the strong
decrease of the chromoelectric field when the tube is collinear with the
magnetic field (case L), while it slightly increases in the TT and TL cases.
Two less prominent but still significant effects that are due to the magnetic
field are the following: \begin{itemize} \item in the longitudinal case the
flux tube gets squeezed, since $E(d, x_t, B)/E(d, x_t, B=0)$ is a decreasing
function of $x_t$, \item in the transverse case the flux tube loses its
cylindrical symmetry, since the results for the cases TT and TL are not equal
to each other.  \end{itemize} This behavior is consistent with the general
picture that emerges from previous studies of static potential and screening
masses \cite{Bonati:2014ksa, Bonati:2016kxj, Bonati:2017uvz}: the magnetic
field acts as transverse confinement catalyser and longitudinal confinement
inhibitor. However previous studies (with the possible exception of
Ref.~\cite{DElia:2015eey}) investigated ``integrated'' quantities (like e.g.
the static potential) and thus could not resolve the difference between the TT
and TL cases.

In the remaining part of this section we will concentrate on the properties of
the flux tube in the L case, i.e. the case in which the separation between the
two color sources is collinear with the magnetic field. In this setup the
cylindrical symmetry of the flux tube is preserved and it is reasonable to
expect the Clem parametrization in Eq.~\eqref{eq:clem_fit} to well describe the
numerical data, which indeed turned out to be the case. 

\begin{figure}
\includegraphics*[width=\columnwidth]{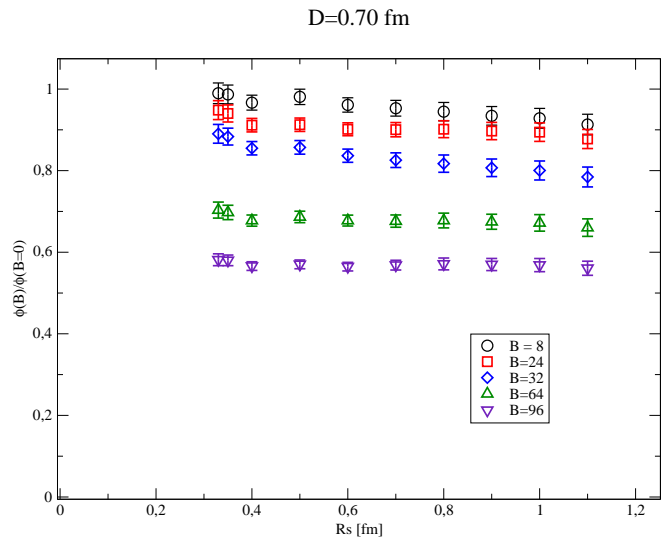}
\caption{Dependence of the ratio $\phi(d, B, R_s)/\phi(d. B=0, R_s)$ on the
smoothing radius $R_s$ for several values of the magnetic field. The figure
refers to the case in which the magnetic field is in the same direction as the
separation between the color sources, and $d\simeq 0.7\mathrm{fm}$.}
\label{fig:phi_fit_ratio}
\end{figure}

The previously noted fact that the ratios $E_l(d, x_t, R_s, B)/E_l(d, x_t, R_s,
B=0)$ are independent of the smoothing radius $R_s$ does not \emph{a priori}
implies the ratios of the fit parameters entering the Clem expression
Eq.~\eqref{eq:clem_fit}, like e.g. $\phi(d, R_s, B)/\phi(d, R_s, B=0)$,
to be also independent of $R_s$. This independence is however numerically
observed to hold true with reasonable accuracy: deviations from a constant do
not exceed the $5\%$ for the range of parameters explored, an example being
shown in Fig.~\ref{fig:phi_fit_ratio}. As a consequence, also the ratios of fit
parameters computed with and without the magnetic field can be used to extract
reliable informations on the effect of $B$ on flux tubes.

To characterize the properties of the flux tube it is convenient to use,
instead of the fit parameters $\phi, \alpha, \mu$, some numerical parameters of
more direct physical and geometrical interpretation. Two such parameters are
the average square width of the flux tube $w^2$ and its energy density per unit
length $\epsilon$, defined by the expressions: 
\begin{equation}\label{w_epsilon}
\begin{aligned}
&w^2=\frac{\int x_t^2 E_{l}(d, x_t)\mathrm{d}^2 x_t}{\int E_l(d, x_t)\mathrm{d}^2 x_t}\\
&\epsilon=\frac{1}{2}\int E_l(d, x)^2\mathrm{d}^2x_t \ .
\end{aligned}
\end{equation}
If we assume for the longitudinal chromoelectric field the expression
in Eq.~\eqref{eq:clem_fit}, using known integrals of the modified Bessel
functions (see, e.g., 
Eq.~5.52.1 and Eq.~5.54.2 in Ref.~\cite{GradshteynRyzhik})
it is not difficult to prove the relations \cite{Cea:2017ocq} 
\begin{equation}\label{w_epsilon_clem}
\begin{aligned}
&w^2=\frac{2\alpha}{\mu}\frac{K_2(\alpha)}{K_1(\alpha)} \\
&\epsilon=\frac{\phi^2\mu^2}{8\pi}\left(1-\frac{K_0(\alpha)^2}{K_1(\alpha)^2}\right)\ ,
\end{aligned}
\end{equation}
that can be used to estimate $w^2$ and $\epsilon$ without having to numerically
perform the integrals on the transverse directions. This is highly desirable
due to the specific form of the integrands of Eq.~\eqref{w_epsilon}: they are
very small everywhere but for a sharp peak at intermediate values of $x_t$ and
this makes the numerical integration unstable with the available numerical
precision.

\begin{figure}[t]
\includegraphics*[width=\columnwidth]{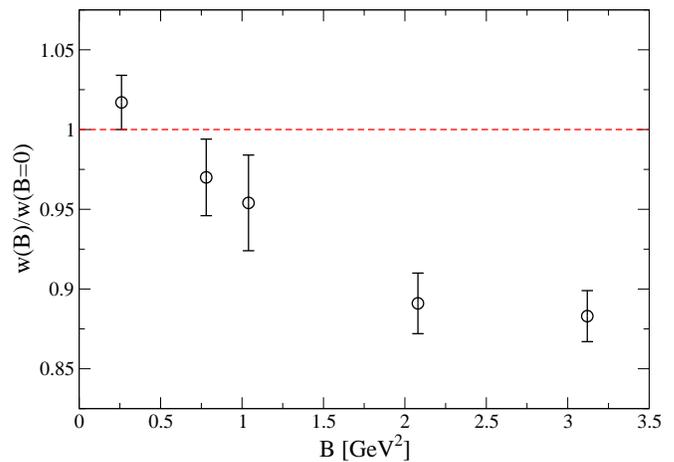}
\caption{The average width of the flux tube as a function of the
magnetic field in the longitudinal configuration (the L case of 
Table~\ref{tab:eq_classes}). Data are normalized to their $B = 0$ values.} 
\label{fig:w_ratio}
\end{figure}

The average square width of the flux tube $w^2$ is not strongly dependent on
the magnetic field and slightly decreases by increasing $B$, being reduced by
about $10\%$ for the largest value of the magnetic field explored, $eB\simeq
3.12~\mathrm{GeV}^2$, see Fig.~\ref{fig:w_ratio}. This is consistent with the
previously noted fact that (in the longitudinal case L) the flux tube gets
squeezed by the magnetic field, see Fig.~\ref{fig:tube_ratio_B}.

Since both the peak value and the square mean width of the flux tube get
smaller in the presence of an external magnetic field, the energy density per
unit length of the tube $\epsilon(B)$ will be a decreasing function of $B$. In
a simple classical picture, the energy density per unit length of the flux tube
is nothing but the string tension and it is thus interesting to compare the
behavior of $\epsilon(B)$ with that of the string tension $\sigma(B)$, that has
been previously investigated in Ref.~\cite{Bonati:2016kxj}. 

\begin{figure}[t]
\includegraphics*[width=\columnwidth]{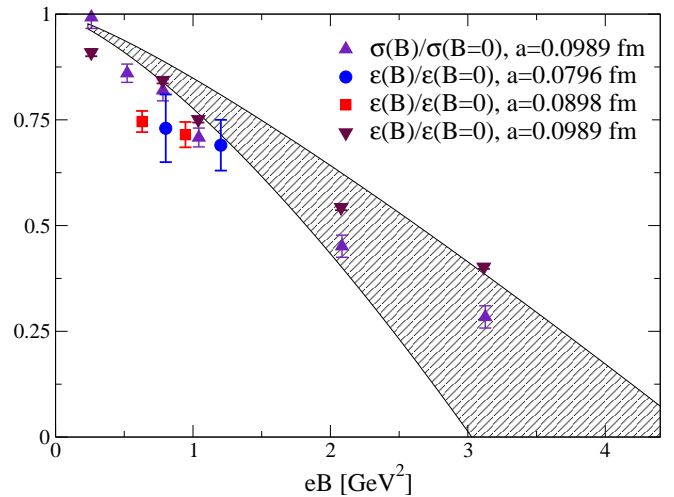}
\caption{Comparison of $\epsilon(d, B)/\epsilon(d, B=0)$ (with $d\simeq
0.7~\mathrm{fm}$) with $\sigma(B)/\sigma(B=0)$ for the longitudinal case L.
The string tension results shown in the picture have been obtained in
\cite{Bonati:2016kxj} using a $48^3\times 96$ lattice with $a\simeq
0.0989~\mathrm{fm}$; the dashed region is the result of the continuum
extrapolation of the string tension performed in \cite{Bonati:2016kxj} assuming
a power-law behavior in $B$.}
\label{fig:string_ratio}
\end{figure}

A direct comparison of $\epsilon(B)$ and $\sigma(B)$ cannot however be
performed, due to the dependence of $\epsilon(B)$ on the smoothing radius
$R_s$.  This dependence disappears in the ratio $\epsilon(B)/\epsilon(B=0)$ and
for this reason we show in Fig.~\ref{fig:string_ratio} the dependence on $B$ of
the ratios $\epsilon(B)/\epsilon(B=0)$ and $\sigma(B)/\sigma(B=0)$ (from
\cite{Bonati:2016kxj}). Taking into account the fact that these two sets of
data have completely different systematics and that in the determination of
$\epsilon(B)$ there is also a theoretical bias (the form
Eq.~\eqref{eq:clem_fit} of the flux tube was explicitly used in
Eq.~\eqref{w_epsilon_clem}), the agreement is reasonable. In future studies it
will be highly desirable to  have more precise data available, in order to
estimate in an unbiased way the energy per unit length of the flux tube.

\section{Conclusions}
\label{concl}

In this work we studied the dependence on a background magnetic field $B$ of
the chromoelectric flux tubes between two static color sources in QCD, using
simulations performed with $N_f=2+1$ dynamical quarks of physical masses.

As a preliminary step we investigated the $B=0$ case, pointing out that the
smoothing procedure used to improve the signal to noise ratio can lead to
significant systematics. We proposed, in alternative to the ``plateau'' method
that is often used in the literature, the ``fixed smoothing scale'' method, in
which the smoothing radius $R_s$ in physical units is kept fixed as the
continuum is approached. Continuum results, however, still display a
significant dependence on $R_s$ and a further extrapolation $R_s\to 0$ is
desirable to obtain physically sensible results,
but our data are not precise enough for this further extrapolation to be
performed reliably.  This makes it very difficult to provide firm results for
interesting quantities like the value of the Ginzburg-Landau parameter $\kappa$
in QCD.

Luckily enough, these problems are absent if one is just interested in studying
the modifications of the flux tube induced by the external magnetic field $B$,
as we showed in Sec.~\ref{resultsnonzerofield}. Since for
$B\neq 0$ rotational invariance is explicitly broken, three different cases
have been studied, corresponding to the inequivalent relative
orientations of the magnetic field, of the Wilson loop and of
the transverse direction.

When the distance between the two static color charges is collinear with the
direction of $B$, the two transverse directions are equivalent, the flux tube
stays cylindrical and both the intensity of the chromoelectric field and the
average square width of the tube decrease with the magnetic field.  When the
static charge separation and the magnetic field are perpendicular, the
cylindrical symmetry of the flux tube is instead broken and the chromoelectric
field inside the flux tube is a growing function of $B$.  

When the flux tube keeps its cylindrical symmetry the tube profile is still
well described by the functional form Eq.~\eqref{eq:clem_fit}, like in the
$B=0$ case. Using this fact we could estimate the energy density per unit
length $\epsilon$ of the flux tube and we verified that, at least at a
semi-quantitative level, the dependence of $\epsilon$ on $B$ is consistent with
the dependence of the string tension $\sigma$ on $B$, as determined previously
in Ref.~\cite{Bonati:2016kxj}.

The general picture that emerges from several studies \cite{Bonati:2014ksa,
Bonati:2016kxj, Bonati:2017uvz}, in which the magnetic field disfavours
confinement in the longitudinal direction and enhances confinement in the
transverse directions, is thus fully consistent with the results of the present
work. Since the properties of the confining potential boil down to properties
of the color flux tube, our results can in fact be seen as the microscopical
origin of the macroscopic effects observed in previous works.

A natural extension of this work would be the study of other field components
of the flux tube: while at vanishing magnetic field the longitudinal
(chromo-)electric field is by far the dominant one, it is conceivable that at
$eB\neq 0$ also other field components could be activated, making the
field structure within the flux tube more complicated.

\acknowledgments

We thank L.~Cosmai, F.~Cuteri, A.~Papa and O.~Pavlovsky for useful discussions.  We
acknowledge PRACE for awarding us access to resource FERMI based in Italy at
CINECA, under project Pra09-2400 - SISMAF. FN acknowledges financial support
from the INFN HPC\_HTC project.  SC acknowledges support from the European
Union's Horizon 2020 research and innovation programme under the Marie Sk\l
odowska-Curie grant agreement No. 642069

\end{document}